# A Practical Marine Wireless Sensor Network Monitoring System Based on LoRa and MQTT

Ao Huang, Mengxing Huang*, Zhentang Shao, Xu Zhang, Di Wu* and Chunjie Cao
State Key Laboratory of Marine Resource Utilization in South China Sea
College of Information Science and Technology, Hainan University
Haikou, China
* Mengxing Huang and Di Wu are the corresponding authors
(Email: huangmx09@163.com, hainuwudi@163.com)

*Abstract*—**Under the advocacy of the international community, more and more research topics have been built around the ocean. This paper proposed an implementation scheme of marine wireless sensor network monitoring system based on LoRa and MQTT. Different from the traditional network architecture, the system was constructed by combining with two network forms, and according to their respective characteristics, the overall design followed the transition from LoRa to MQTT. We first used LoRa to interconnect the sensor nodes with the gateway, and on this basis, the collected data was sent to the server visualization platform through MQTT, the backend management server would continuously refresh the monitoring page. At the same time, the client could use a browser-based web application to directly access and call data for global maritime information monitoring. In the future, we will further improve the system and optimize the algorithm, to achieve more dimensions and deeper exploration of the underwater world.**

## I. INTRODUCTION

In recent years, people are paying more and more attention to the utilization and development of marine resources. For example, in order to fully exploit the role of the ocean in predicting global climate change, in 1998, a large ocean observation program, Argo, was launched by atmospheric and marine scientists in the United States and other countries [1]. The marine environment is complex and changeable. At the same time, traditional monitoring methods have many loopholes, such as long data collection period, difficult data organization, and backward information visualization methods, which makes it difficult to restore the real marine ecological environment. With the advancement of information technology, modern marine environments monitoring technology in the world is gradually developing towards high timeliness, high integration, intelligence and internet [2]. Wireless sensor networks [3], due to their low cost and convenient deployment, have shown broad application prospects in the field of marine environmental information exploration.

In view of the existing single-network transmission data method, this paper proposes a construction method of a multi-modal network combined maritime monitoring system--an implementation scheme of marine wireless sensor network monitoring system based on LoRa and MQTT [4]. The system consists of three parts: sensor node, gateway and server platform. A temperature sensor, a main control chip, and a LoRa transceiver module are installed inside the sensor node. Its overall package has made the necessary waterproofing measures and GPS positioning. As for the gateway, we use ESP32 as the main control chip to complete the conversion of the network protocol from LoRa to Wi-Fi. In addition, we have created a server platform to visualize data information. During system operation, we first use multiple sensor nodes deployed in or near the monitoring area to collect underwater temperature information. Then, the data is transmitted to the gateway based on LoRa, and the Wi-Fi module of the ESP32 on the gateway is used to upload the data to the cloud server of the Internet through the MQTT protocol, which is convenient for the client to access and call the resulting data information at any time. This paper first gives a summary design of the system, then describes the detailed design and implementation of each sub-module, and gives the results of relevant experimental verification. At the end of the paper, we have proposed a further improvement plan for the entire system.

## II. SYSTEM ARCHITECTURE

This set of wireless sensor network monitoring system combines LoRa and MQTT to collect and manage the temperature of the detected seawater. Fig. 1 shows us the overall block diagram of the system. The system is mainly composed of three units, including data acquisition, wireless data transmission and terminal monitoring.

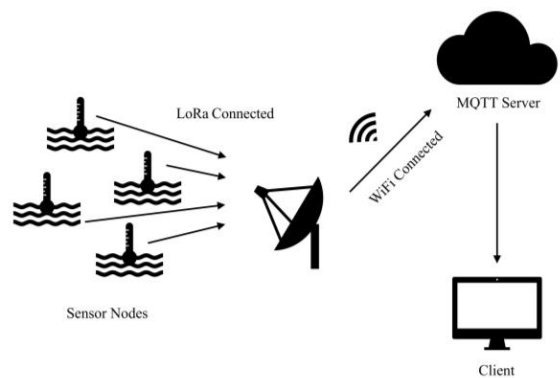

Figure 1. The architecture of the system

## III. SYSTEM HARDWARE DESIGN

In order to complete the monitoring and collection tasks of maritime data, we first need to realize the communication between the sensor nodes and the information feedback platform. Facing the complex natural environment at sea, the 433M frequency band has a longer transmission distance and higher stability [5]. We compared the SX1278 chip with the HC-12 chip that both works at 433MHz. The key parameters of the two chips are shown in table1. We found that the SX1278 chip has better overall performance, its transmission distance can be greatly extended, and the system design is simplified without the need for a relay device and a complicated communication infrastructure. SX1278 has obvious advantages in anti-blocking and selectivity due to the spread spectrum communication technology [6] and greatly improves the anti-interference tolerance of the system, making the data transmission more stable. In addition, SX1278 solves the fatal flaw in the power consumption of HC-12, which minimizes current consumption. In sleep mode, the power consumption of the chip is negligible. Therefore, we chose the SX1278 chip for the communication design of the system.

TABLE I. SX1278 AND HC12 KEY PARAMETERS

| Chip model | Frequency range (MHz) | Est.sensitivity (dBm) | Communication distance(m) |
|---|---|---|---|
| SX1278 | 137 ~ 525 | -132 | 5000 |
| HC-12 | 433 ~ 473 | -117 | 600 |

### A. Sensor node design

Due to the difficulty of offshore operations and the wide monitoring area, designing a simple, accurate and low-cost wireless sensor node is the key to this system. The node integrates a temperature sensor, ATmega2560, LoRa wireless transceiver module, GPS locator and power module, its hardware block diagram and its main controller circuit schematic are shown below (Fig. 2 and Fig. 3).

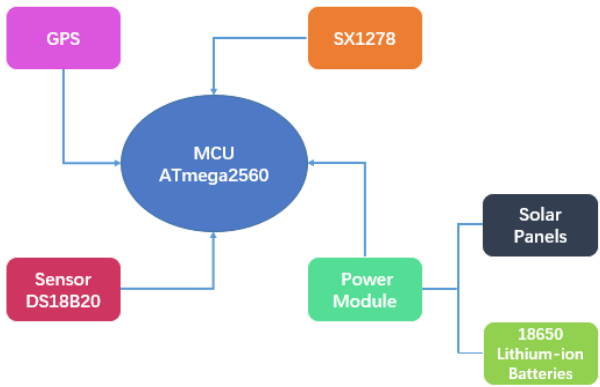

Figure 2. Node hardware block diagram.

The RF chip sx1278 lead to the SPI interface. The node master controller use ATmega2560 to communicate with the sx1278 chip through SPI, to initialize the chip, configure communication parameters, switch the working mode, and send and receive data. In this system, we used the DS18B20 temperature sensor to collect temperature information. The DS18B20 is a digital single-bus intelligent temperature sensor manufactured by DALLAS Semiconductor Corporation of the United States. Unlike conventional thermistors, it is capable of directly reading the measured temperature. According to the actual requirements, we designed a 4.7 kΩ pull-up resistor in parallel with the two leads of the DS18B20, and connected to the main controller for temperature measurement. By programming we have implemented multiple numerical readings that was simple and intuitive. The node used a direct power supply mode of four 18650 lithium-ion batteries. It was designed with extremely low power requirements, which significantly extended the battery power supply time. After the node is recycled, we can also recharge it to save costs. Further, we have attached 6 solar panels on the inner wall of the node to reduce the working pressure of the lithium battery, so that we can maximize the life of the node. Using the positioning function of the GPS module, it is easy to grasp the exact location information of the monitoring sea area, thus helping us to establish a more complete model of the marine physical environment.

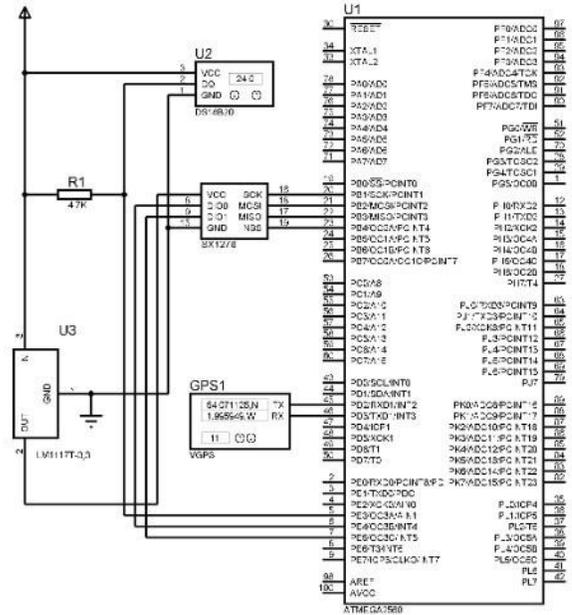

Figure 3. Main controller circuit schematic

### B. Node overall package

The entire node is required to be placed in seawater, so its water resistance and corrosion resistance are extremely high. We used a 110mm outer diameter and 104mm inner diameter acrylic tube for encapsulation. The overall shape was cylindrical. Fig. 4 is a specific package effect diagram of the node. Its two ends were sealed with silicone plugs. All the interfaces are treated with multi-layer glass glue. Even if it was immersed in seawater for a long time, the internal sensor nodes could be basically guaranteed to be normal and effective. In order that the node may be delivered to the water just floating on the water surface to maintain balance, according to mechanics, we installed four lithium-ion batteries on the inner wall of the tube that sunk in the water to increase its weight. The tube body leaded out a small hole with a diameter of about 18mm, after inserting the SP13-3 core waterproof aviation plug, we connected the three leads of the DS18B20 to the copper core inside the plug, and then tightened the bolt to fix the sensor. The other side of the tube was affixed with a row of solar panels to maximize the area of the light, ensuring that the nodes can use the light to capture a portion of the energy. In addition, we used the ATmega2560 expansion board to integrate the cable, which was glued to the tube wall with chips such as GPS by using glass glue. The division of labor within the entire node was clear and modularized. There was a waterproof button on the outside, and the button could be pressed to start the node to work normally.

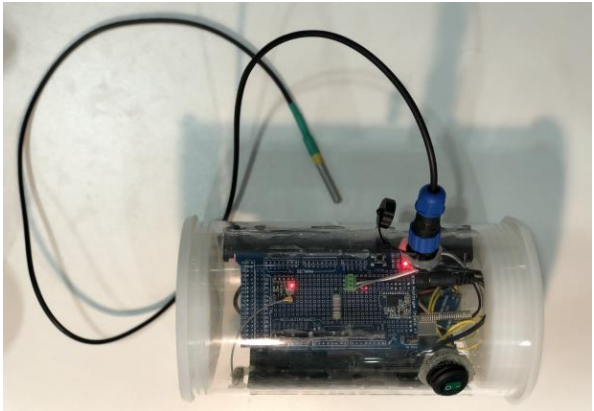

Figure 4. Node overall package

## IV. SYSTEM NETWORK SETTINGS

Because of the limitations of a single sensor, we used a wireless sensor network to construct the system, it has distributed characteristics compared to single point sensor measurement methods [7]. In order to reduce the accuracy requirements of individual sensors and improve the fault tolerance of the system, we deploy sensor nodes on a large scale at sea to understand the environmental conditions of the entire waters from different spatial perspectives, the spatial correlation of the information collected by the nodes is used to obtain more accurate environmental information. In addition, building a physical monitoring system in this way can increase the water coverage to a large extent and reduce the blind area of monitoring.

In terms of network settings, we use a combination of two network protocols, including LoRa and MQTT. LoRa [8] is a radio frequency modulation and demodulation technology based on spread spectrum technology developed by Semtech, which provides users with ultra long distance, low power consumption and safe data transmission. MQTT [9] is an instant messaging protocol developed by IBM that is designed for communication with remote sensors and control devices that have limited computing power and operate on low-bandwidth, unreliable networks.

The sensor node network is built using LoRa. After completing the collection of related data, the Wi-Fi and Internet are used to upload data to the server. Previously, only those high-level industrial radio communications would incorporate these technologies, but in this system, we also use a combination of LoRa and Wi-Fi networks to achieve the best communication results. First, we introduce LoRa to configure wireless sensor networks to complete data acquisition and transmission of maritime information. Moreover, we use ESP32 as the core chip of the gateway, the design of the gateway is shown in the Fig. 5 Equipped with LoRa transceiver module, gateway and sensor nodes are connected in a star network, then use IEEE 802.11 b/g/n based Wi-Fi to connect to the server. We use MQTT to communicate with the Internet, and finally upload the data to our self-built server platform through the Internet to dynamically display the collected real-time data information and its curve. Besides, the RESTful API is open for remote access, data downloads and calls.

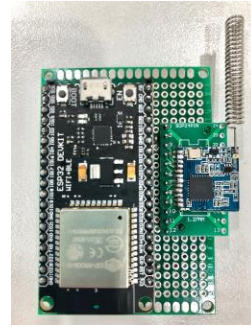

Figure 5. The design of the gateway

## V. SYSTEM EXPERIMENT RESULTS AND ANALYSIS

According to the rules and methods of design, we conducted multiple experiments on land and under water at different times and locations.

### A. Program feasibility assessment test

In the early days of system design, we evaluated the feasibility of combining LoRa and Wi-Fi communication by trying to communicate in different land environments. The results showed that in the open area, the coverage of the combined network can reach 3 to 4 kilometers. In urban areas where buildings are densely populated, its communication capabilities have declined. However, within the transmission range of 1.5 km, the reliability of data transmission could still be trusted. Except for occasional data packet loss, no large transmission problem was found. Those experiments proved that the scheme of data transmission in combination network can be adopted.

### B. System operation effect measurement

After completing the package of the entire system, we have conducted several launch experiments, take Fig. 6 as an example, and its satellite renderings are shown in Fig. 7. This system aimed to achieve the collection and monitoring of sea temperature information. The sensor node, the gateway node and the server platform were respectively arranged in corresponding positions according to requirements, and each node was powered on and the system was debugged. After the debugging was completed, the sensor node joined the wireless network, automatically connect to our Wi-Fi router, the node would try to log on to the MQTT server and confirm that whether the connection was successful. If the connection is successful, the ESP32 chip will acquire sensor data and build a JSON format for uploading. The temperature data continuously be sent to the monitoring platform, users can communicate with the data control terminal via the Internet.

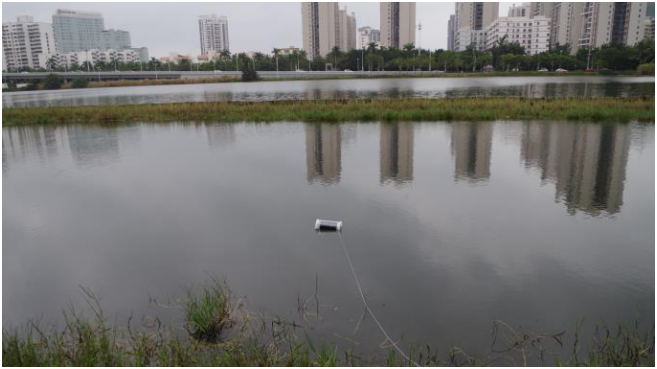

Figure 6.　Outdoor water environment test

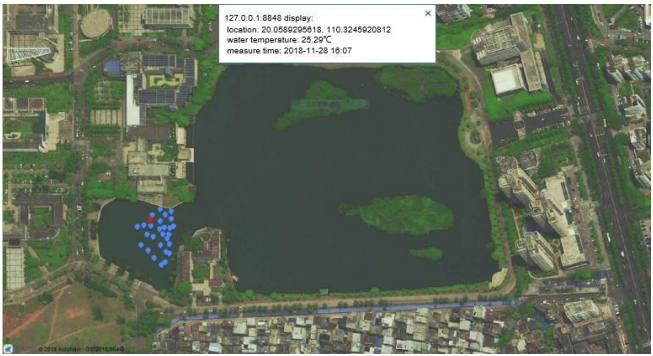

Figure 7.　Satellite renderings

MQTT is a line layer protocol that any client implementing the protocol can connect to Apollo. After the login is successful, the console page is as shown below in Fig. 8. The system's processing of data will be displayed on the system interface, and the data processing interface was periodically refreshed in order to ensure the real-time nature of the data. There is a running interface is shown in Fig. 9.

The measurement results had slight fluctuations and errors, which could be compensated by the software in the upper computer. The data processing interface refresh frequency of the system was 10ps, and the data displayed on the interface could directly analyze the environmental conditions at sea and complete the monitoring task.

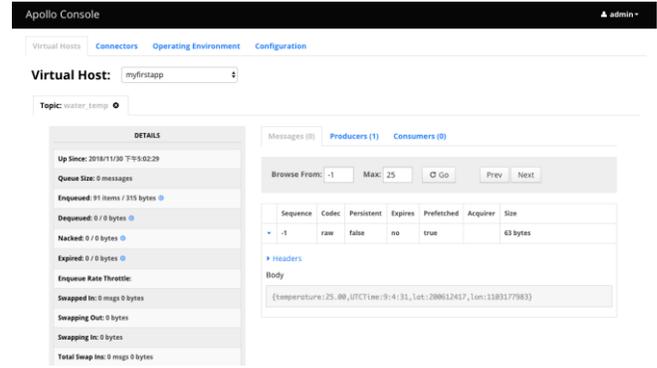

Figure 8.　Console page of Apollo

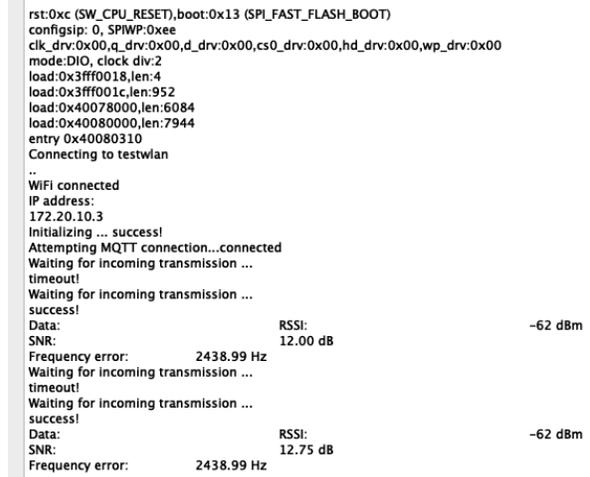

Figure 9.　Serial port monitoring

When the wireless node system is used outdoors, changes in weather conditions in the field will also affect the transmission of wireless signals. The main meteorological conditions considered in the wild are temperature and humidity changes. It has been verified by experiments that the influence of temperature and humidity conditions on wireless signal transmission is irregular, but the effects are not obvious, methods such as mean or weighted before and after measurements eliminate their effects. At present, we can obtain the RSSI value of the wireless signal through the program interface, and collect the RSSI value at different communication distances in units of 60 meters, by quantifying the relationship between the two, we can obtain the fitting curve (Fig. 10).

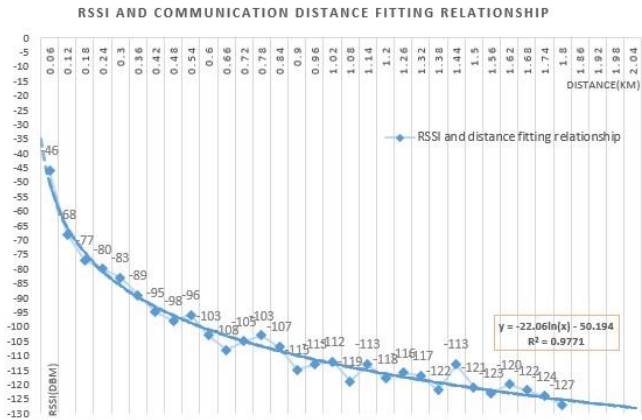

Figure 10. RSSI and distance fitting relationship

The curve shown in Fig. 10 has a fit of 0.9771, which approximately satisfies the mathematical expression:

$$y = -22.06 \ln(x) - 50.194$$

It indicates that there is a certain functional relationship between the RSSI value and the wireless signal transmission distance. The RSSI measurement is reproducible and interchangeable, and its modest changes are regularly queried in the application environment. From the practical experience: For the outdoor water environment, when the whole set of equipment is within a certain coverage of the wireless signal, the RSSI value meets the basic communication requirements, at the same time, the collected sensor data also has high reference value and practical significance.

## VI. CONCLUSION AND FUTURE DIRECTIONS

In this paper, we discussed the implementation of a wireless sensor network for collecting seawater temperatures. Over a period of time, we have focused our research on the innovation of data communication models. After measuring the temperature with the sensor node and uploading the data to the server platform with the LoRa and MQTT combination network, the users can directly access the monitoring result by logging in to the WEB. However, due to the acquisition of the data source was only the temperature information, our monitoring of the overall seawater quality was not enough. In the future, we plan to add more sensors such as pH sensors, oxygen concentration sensors and chlorophyll sensors inside the nodes, and install underwater probes to explore the underwater biological world, so that we can establish a more complete marine ecosystem monitoring platform. At the same time, optimizing the algorithm to further improve the accuracy and stability of data uploading, this will also become an important part of our research in marine topics.

ACKNOWLEDGMENT

This research received financial support from the Natural Science Foundation of Hainan province (Grant #:617062), National Natural Science Foundation of China (Grant #: 61462022), Major Science and Technology Project of Hainan province (Grant #: ZDKJ2016015).